# Access-Point to Access-Point Connectivity for PON-based OWC Spine and Leaf Data Centre Architecture

**Abrar S. Alhazmi, Sanaa H. Mohamed, Ahmad Qidan, T. E. H. El-Gorashi, and Jaafar M. H. Elmirghani**
Department of Engineering, Kings College London
Abrar.alhazmi@kcl.ac.uk, sanaa.mohamed@kcl.ac.uk, ahmad.qidan@kcl.ac.uk, taisir.elgorashi@kcl.ac.uk
jaafar.elmirghani@kcl.ac.uk

**ABSTRACT**

In this paper, we propose incorporating Optical Wireless Communication (OWC) and Passive Optical Network (PON) technologies into next generation spine-and-leaf Data Centre Networks (DCNs). In this work, OWC systems are used to connect the Data Centre (DC) racks through Wavelength Division Multiplexing (WDM) Infrared (IR) transceivers. The transceivers are placed on top of the racks and at distributed Access Points (APs) in the ceiling. Each transceiver on a rack is connected to a leaf switch that connects the servers within the rack. We replace the spine switches by Optical Line Terminal (OLT) and Network Interface Cards (NIC) in the APs to achieve the desired connectivity. We benchmark the power consumption of the proposed OWC-PON-based spine-and-leaf DC against traditional spine-and-leaf DC and report 46% reduction in the power consumption when considering eight racks.

**Keywords** Passive optical networks (PONs), Optical Wireless Communication (OWC), Data Centre (DC), Network Interface Card (NIC), and Optical Line Terminal (OLT).

## 1. INTRODUCTION

The rapid growth of big data and cloud computing has led to a significant increase in traffic within Data Centres (DC). Most of this traffic is transferred between servers in racks, creating challenges for Data Centre Network (DCN) architectures [1], [2] and [3]. With hundreds of thousands of data storage and processing servers, it is essential to design DCNs that meet the high bandwidth, flexibility, scalability, and energy efficiency requirements [4], [5]. Due to flexibility, cabling, cooling, and power consumption limitations, traditional wired DCN architectures cannot efficiently support these requirements [6], [7]. Introducing Optical Wireless Communication (OWC) solutions into DCs could be a promising solution that can overcome the limitations of traditional DCs. OWC in DCs reduces cabling requirements and enables flexibility in the designs as the configuration of the transceivers can be performed without the need for complex rewiring. Furthermore, OWC in DCs consumes less power and reduces equipment size, and hence, reduces cooling requirements, while providing high data rates [8-20]. Passive Optical Networks (PONs) have also been considered for DCs as they offer high bandwidth, flexibility, cost-effectiveness, and energy efficiency due to the use of passive components [21-23]. Combining PONs with OWC in DCs can offer numerous benefits particularly in terms of increased energy efficiency and flexibility.

In this work, we propose the use of a PON DC design introduced in [21] and [23] that utilise Network Interface Cards (NIC) in the servers, in an OWC-based spine and leaf DC architecture [24-26]. The OWC-based DC architecture connects the racks using Wavelength Division Multiplexing (WDM) Infrared (IR) transceivers that are placed on top of the racks and at distributed APs in the ceiling. Each transceiver on a rack is connected to a leaf switch that connects the servers within the rack. The PON design is used to achieve AP-to-AP connectivity, where each AP is supported by a NIC. The AP-to-AP PON, in addition to an Optical Line Terminal (OLT) are used to replace the spine switches for the connectivity between the racks and the connectivity with external networks.

The rest of this paper is organised as follows: Section 2, describes the proposed AP-to-AP based OWC-PON DC architecture. Section 3 presents a benchmark study that compares the power consumption of our proposed architecture with that of traditional spine and leaf data centre architecture. Finally, Section 4, concludes the paper and describes areas of future work.

## 2. THE PROPOSED AP-TO-AP BASED OWC PON SPINE-AND-LEAF DC ARCHITECTURE

The AP-to-AP-based design enables DCs to support high bandwidth and low latency applications. It requires a NIC per AP that enables connectivity between the APs and between the APs and the OLT [21], [22] and [23]. The architecture is divided into two layers as shown in Figure 1. The first layer represents the OWC layer, which is discussed in our previous work [24], [25] and [26], while the second layer is the AP-to-AP PON layer, where eight APs are used to support 8 racks. In the PON layer, APs are divided into two groups, where each group contains four APs. The architecture provides direct AP-to-AP connections for inter-group communication.

We consider two optical switches (one in each group) for intra-group connections, where the optical switch in each group is used to convey traffic between APs in the same group. There is also a direct connection between one of the APs in each group and the OLT. This network architecture shares similarities with the network design proposed in [23]. The work in [23], proposed an optical backplane to connect entities within the same group. However, in our proposed architecture, the optical backplane is replaced by the optical switch. This is mainly due to the advantages that can be attained by adopting optical switches such as the commercial availability of optical switches such as Micro-Electro-Mechanical Systems (MEMS) switches with low latency and good scalability [27]. The NICs enable forwarding the traffic between the APs in the same group via the optical switch, APs in different groups via the optical switches and the OLT, and between the APs and the OLT via the optical switch. The OLT controls all communication in this architecture, select the paths, and can relay traffic between different groups. Also, the OLT is responsible for the communication with external network layers.

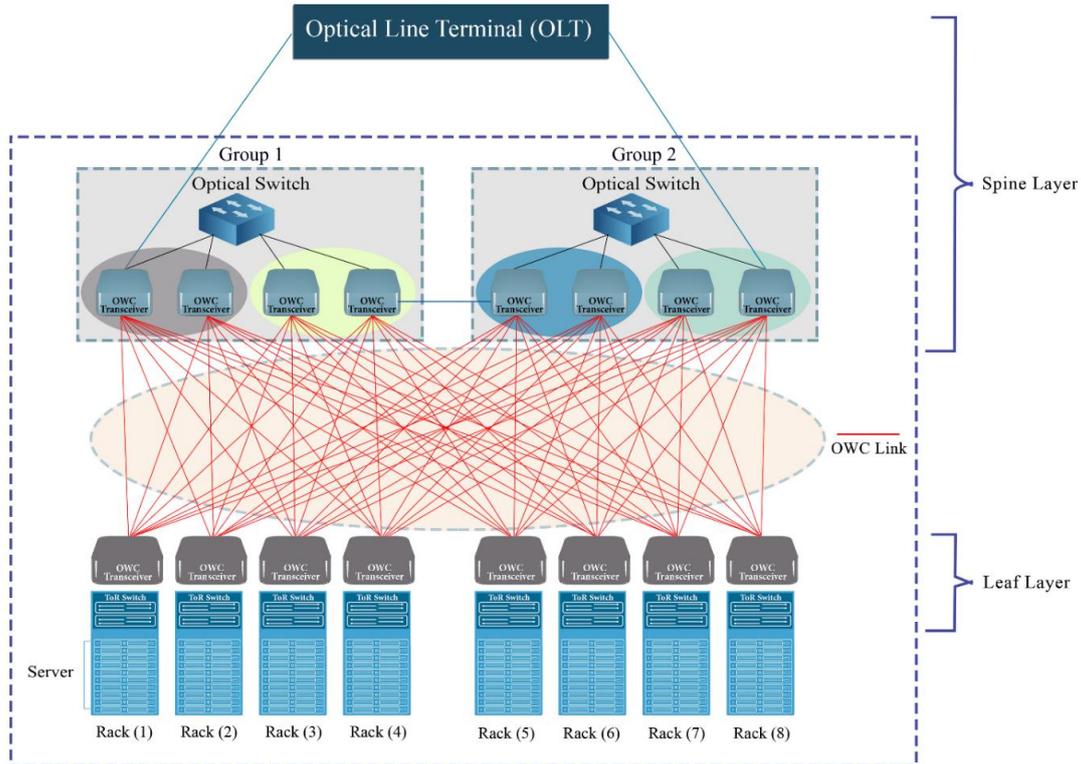

**Figure 1: The proposed AP-to-AP OWC spine and leaf DC architecture.**

### 3. POWER CONSUMPTION BENCHMARK

In this section, we present a benchmarking study for the power consumption of our proposed designs in comparison with traditional spine and lead DC architecture when considering 8 racks and 8 servers per rack in the two architectures (i.e., the traditional spine and leaf, and the OWC-PON-based spine and leaf DC with the AP-to-AP connectivity). The power consumption values of different devices in the traditional spine and lead DC architecture are listed in Table 1. The power consumption values of different devices in our proposed OWC-PON-based spine and leaf DCs with the AP-to-AP PON are listed in Table 2. Equation (1) calculates the power consumption of a spine and leaf DC network architecture:

$$P = (Ps\ Ns) + (Pl\ Nl) + (Pcs\ Ncs), \qquad (1)$$

**Table 1 : Power consumption of different devices in traditional spine and leaf DCs**

| Network device | Power consumption (W) |
|---|---|
| Spine Switch | 660 [28] |
| Leaf Switch | 508 [28] |
| Server's Transceiver | 3 [28] |

where $Ps$ is the power consumption of a spine switch, $Ns$ is the number of spine switches used in the architecture, $Pl$ is the power consumption of a leaf switch, $Nl$ is the number of used leaf switches in the architecture, $Pcs$ is the power consumption of a server's transceiver, and $Ncs$ is the number of transceivers used in the architecture.

Equation (2) calculates the power consumption of the proposed OWC-PON-based spine and leaf DC with AP-to-AP PON:

$$P = (Po\ No\ ) + K + T\ Nm + (O\ Nk) + (Pl\ Nl) + (Pcs\ Ncs), \qquad (2)$$

where $Po$ is the power consumption of the OWC transceivers, $No$ is the number of the OWC transceivers, $K$ is the power consumption of the OLT, $T$ is the power consumption of the optical switches, $Nm$ is the number of the optical switches, $O$ is the power consumption of the NIC, $Nk$ is the number of the NICs. Please note that we omit the calculation of the power consumption of the OWC transceivers for both architectures.

**Table 2 : Power consumption of different devices in the proposed OWC-PON-based spine and leaf DC architectures AP-to-AP based backhaul.**

| Network device | Power consumption (W) |
|---|---|
| OWC transceiver | 0.4 [29] |
| Leaf Switch | 508 [28] |
| OLT | 480 [30] |
| Server's Transceiver | 3 [28] |
| NIC | 45 [31] |
| Optical Switch | 75 [27] |

We consider eight racks with total of 64 servers. For the traditional spine and leaf DC architecture, we considered eight spine switches and eight leaf switches. The number of NICs in the AP-to-AP architecture is equivalent to the number of servers. As illustrated in Figure 2, our proposed the AP-to-AP based architecture minimises the total networking power consumption by about 46% compared to traditional spine and leaf DC. The reason for the decrease in power consumed is due to the replacement of the spine layer with PON components. This demonstrates how our proposed architecture is efficient in lowering the power consumption. For businesses aiming to maximise power efficiency without sacrificing performance, this architecture offers a highly feasible solution.

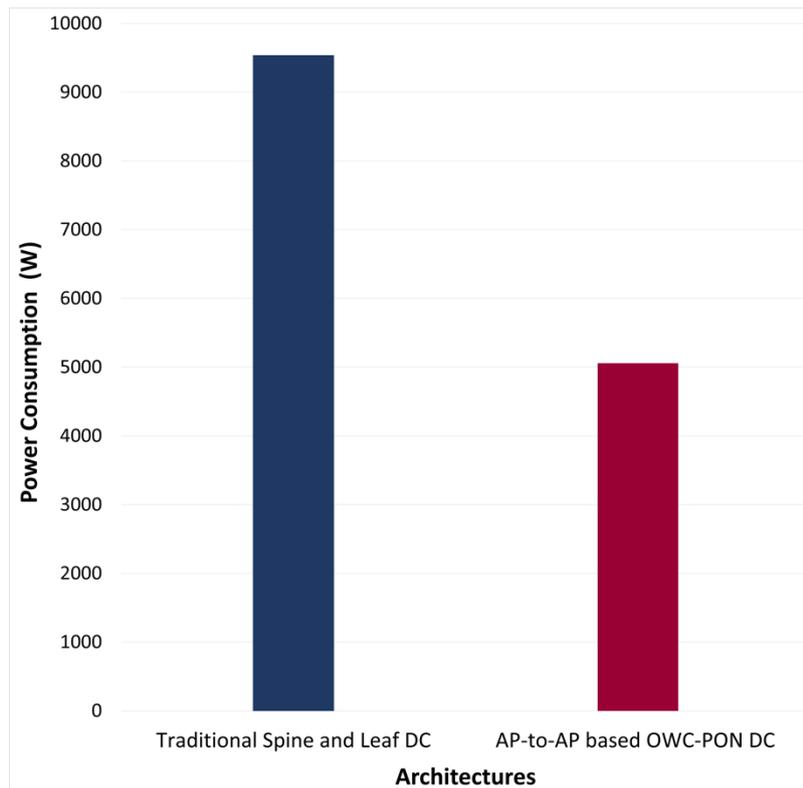

Figure 2 : Power consumption benchmarking.

### 4. CONCLUSIONS

In this paper, we proposed an AP-to-AP based PON connectivity for the OWC-based spine and leaf DC. We replace the spine switches by an OLT, in addition to the AP-to-AP PON to achieve connectivity in the DC. We maintained the leaf switches to handle intra-rack connectivity. The AP-to-AP PON enable connecting the APs together and provides connectivity with the OLT. The power consumption of the OWC-PON-based DC with the AP-to-AP based PON was compared to the power consumption of traditional spine and leaf DCs and reduction by about 46% is achieved. Future work includes experimental evaluations for the proposed OWC-PON-based DC and evaluation of the impact of the OWC links on workload placement.

### ACKNOWLEDGEMENTS

The authors would like to acknowledge funding from the Engineering and Physical Sciences Research Council (EPSRC) INTERNET (EP/H040536/1), STAR (EP/K016873/1) and TOWS (EP/S016570/1) and TITAN (EP/X04047X/1 and EP/Y037243/1) projects. Abrar would like to thank Taibah University in the Kingdom of Saudi Arabia for funding her PhD scholarship. All data are provided in full in the results section of this paper.